\documentclass[useAMS,usenatbib]{mn2e}

\def\simlt{\lower.5ex\hbox{$\; \buildrel < \over \sim \;$}}
\def\simgt{\lower.5ex\hbox{$\; \buildrel > \over \sim \;$}}
\def\etal{{\it et al.}}

\def\ie{{\it i.e.}}
\def\eg{{\it e.g.}}
\def\cf{{\it c.f.}}
\usepackage{graphicx}

\DeclareGraphicsExtensions{.eps,.ps}

\title[Charge Transfer Inefficiency in HST since SM4]{Charge Transfer Inefficiency in the Hubble Space Telescope since Servicing Mission 4}
\author[R. Massey]{Richard Massey\\Royal Observatory, Blackford Hill, Edinburgh EH9 3HJ\vspace{-1mm}}
\begin{document}

\date{Accepted ---. Received ---; in original form ---}

\pagerange{\pageref{firstpage}--\pageref{lastpage}} \pubyear{2010}

\maketitle

\label{firstpage}

\begin{abstract}

We update a physically-motivated model of radiation damage in the {\em Hubble
Space Telescope} Advanced Camera for Surveys/Wide Field Channel, using data up
to mid 2010. We find that Charge Transfer Inefficiency increased dramatically
before shuttle Servicing Mission 4, with $\sim1.3$ charge traps now present per
pixel. During detector readout, charge traps spuriously drag electrons behind
all astronomical sources, degrading image quality in a way that affects object
photometry, astrometry and morphology. Our detector readout model is robust to
changes in operating temperature and background level, and can be used to
iteratively {\it remove} the trailing by pushing electrons back to where they
belong. The result is data taken in mid-2010 that recovers the quality of
imaging obtained within the first six months of orbital operations.

\end{abstract}


\section{Introduction}

The harsh radiation environment above the Earth's atmosphere gradually degrades
all electronic equipment, including the sensitive  Charge-Coupled Device (CCD)
imaging detectors used in the {\em Hubble Space Telescope} (HST) Advanced Camera
for Surveys (ACS)/Wide Field Channel (WFC). The detectors work by collecting
photoelectrons in a potential well at each pixel. At the end of an exposure,
these electrons are then transferred, row by row, to an amplifier at the edge of
the device, where they are counted. However, radiation damage to the silicon
lattice creates charge traps that temporarily capture electrons and release them
only after a characteristic delay. Any electrons captured during the transfer to
the readout register can reemerge, several pixels later, as a spurious ``Charge
Transfer Inefficiency'' (CTI) trail behind every bright source.

CTI trailing is particularly troublesome because the amount of flux trailed is a
nonlinear function of the flux, size and shape of a source. The effect
is therefore {\it not} a convolution. 
A multitude of ad-hoc schemes have been invented to estimate (and subtract) the
effect of CTI from catalogues of object photometry, astrometry and shape
\citep[\eg][]{chiaberge,cawley02,rhodes07}. However, since CTI moves electrons
around fairly predictably at the image level, the ideal approach for correction
is to directly shuffle those electrons back to where they belong. Since detector
readout is the last process to happen during data acquisition, this can be
conveniently carried out as the first process in data processing. While no
algorithm can undo the nonlinear movement of electrons in a single step,
\cite{bristow03im} and \cite{piatek05} proposed an iterative algorithm to remove
trailing by repeatedly (re-)adding new trailing. This requires a model of the
(forward) readout process. A physically-motivated model was developed for
ACS/WFC by \cite{m10}, using measurements from trails behind warm pixels in
science imaging.

This letter updates the \cite{m10} CCD readout model and pixel-based CTI
correction. In \S\ref{sec:method}, we account for an additional species of
charge trap with a long characteristic release time, and measure the density of
traps in the detector up to mid-2010. In \S\ref{sec:implementation}, we
implement the improved CTI correction algorithm and evaluate its performance. In
\S\ref{sec:conclusions}, we discuss the overall performance of the detectors, in
light of changes to their operational temperature and the long period during
which they were offline.

\section{Updated CCD readout model} \label{sec:method}

\subsection{Well filling rate}

We use a ``volume-driven'' CCD readout model, whose first ingredient is the rate
at which electrons fill up the potential well in a pixel. A cloud of $n_e$
electrons grows in size as electrons are added, and a cloud with a larger
cross-sectional area will be exposed to more charge traps when it is swept
through the silicon lattice during readout. As first suggested by
\citet{biretta05}, if the traps are uniformly distributed in 3D, the well
filling rate can be measured using hot pixels (which would appear as isolated
$\delta$-functions in the absence of radiation damage) in ordinary, on-orbit
imaging -- from the increasing fraction of electrons trailed behind increasingly
warm pixels. Universally, the fraction of trailed electrons is greatest for
faint sources, demonstrating that the size of the cloud grows more slowly than
the number of electrons. This crucial point explains why CTI is nonlinear.

\cite{m10} parameterised the height $h(n_e)$ of a cloud as zero for the first
$d\approx100$~electrons, then increasing as $h\propto(n_e-d)^\alpha$ where
$\alpha\approx 0.58$. The first electrons were assumed to reside in a
supplementary buried channel or ``notch'' specifically intended to compress
their volume and minimise the degradation of very faint sources. The notch is
created by doping the silicon lattice, but HST engineers now believe that the
initial atomic implant in the ACS/WFC detectors was unstable and has diffused
(Linda Smith, priv.\  comm.\ 2010). If this were true, it would result in
trailing behind even faint sources. The interpretation of our (updated)
measurements of faint in science images is hindered by the zodiacal sky
background, but the data are consistent with no notch. \citet{a10} studied warm
pixels in dark exposures, which have less noise, and confirmed trails behind
warm pixels containing as few as $\sim$20~electrons. We shall therefore adopt a
model in which the notch is no longer operational $d\equiv 0$. Strictly, we
should model the gradual disappearance of the notch over time -- but the same
sky background that makes it difficult to measure this effect also hides real
science data from its influence.

\begin{figure}
 \includegraphics[width=84mm]{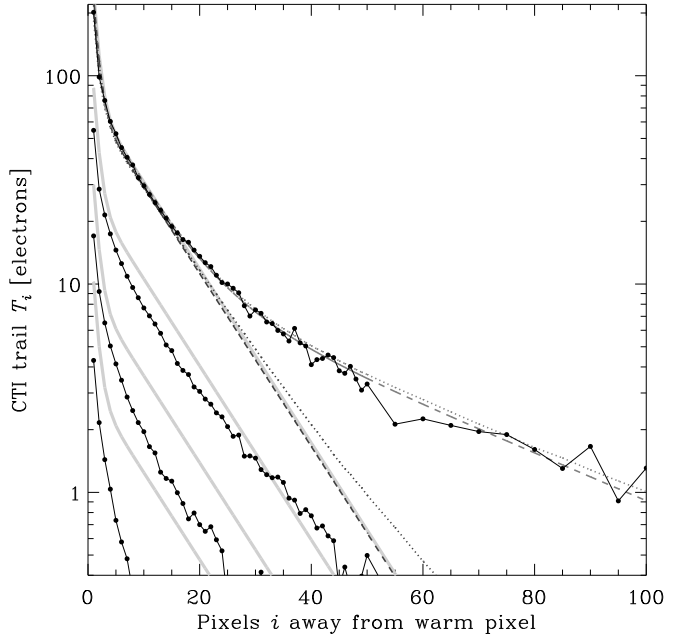}
 \caption{CTI trailing behind warm pixels in dark exposures. 
 Black points reproduce the measurements from figure~5 of \citet{a10}, for warm pixels at least 1500 pixels from the readout
 register and containing approximately 20, 200, 2000 and 20000 electrons (bottom to top).
 Solid grey lines show predictions of the \citet{m10} model, which used only the first 9 pixels behind warm pixels, and has had to be extrapolated down to the lower trails.
 The agreement between these completely independent analyses is impressive.
 Dashed lines show best-fit models of the trail behind the brightest curve (which is most reliably measured), using
 a double exponential \citep[with the decay times of][]{m10} and a triple exponential with free parameters.
 The dotted lines add secondary electron capture to the calculation, whereby trailed electrons can be recaptured and retrailed.
 This represents the difference between our full algorithm and the (much faster) approximation of \citet{a10}. \label{fig:trail}}
\end{figure}

The profiles of CTI trails in dark exposures are reproduced from \cite{a10} in
figure~\ref{fig:trail}. The relative fraction of electrons trailed behind
increasingly warm pixels confirms that $\alpha\approx 0.57$. However, when
ignoring a notch, measurements from science images like those in \citet{m10}
(see \S\ref{sec:measurements}) prefer $\alpha=0.465\pm 0.016$ both before and
after Servicing Mission 4. The effect of this parameter is apparent as the
difference between the data and the solid grey curves of figure~\ref{fig:trail},
which show the \citet{m10} prediction but using the lower value of $\alpha$. For
very hot pixels, the first $\sim$10~pixels of the predicted trail are within 3\%
of the \citet{a10} data (\cf\ the best-fit dashed curve) -- an impressive
agreement considering these measurements are completely independent. However,
predictions of the relative trail heights begin to disagree when the model is
extrapolated down towards faint trails. Since the measurements of these faint
trails are affected by a complex interaction with the sky background, we adopt
our measurement of $\alpha=0.465$ because its origin is closest to the data we
will eventually want to correct.

\subsection{Charge trap species}

The second ingredient of a CCD readout model is the density and characteristic
release times of charge traps. Shockley-Read \& Hall theory of solid-state
devices \citep[\eg][]{hardy98,janesick01} suggests that we can expect several
distinct species of traps at a variety of energies $\Delta E$ below the band
gap, all of which capture charge almost instantly then release it with a
probabilistic delay governed by an exponential $e^{-t/\tau}$. The characteristic
release time $\tau$ depends upon operating temperature as $\tau\propto
T^{-2}e^{\Delta E/kT}$.

In early ACS data, \citet{m10} found two species of traps with characteristic
release times $\tau=\{0.88,10.4\}$ multiplied by the $3212\mu$s CCD clocking
speed, and associated them with impure E-centre complexes at $\Delta E=0.31$ and
0.34~eV \citep{hopkinson01}. The trap species were present in a density ratio of
$1:3$. 

In July 2006, the operating temperature of the WFC detectors was lowered from
-77C to -81C \citep{mack06,mack07}.
\citet{a10} modelled the trail profiles in subsequent imaging using an empirical look-up
table. However, the smooth curves overlaid on their data in figure~\ref{fig:trail}
demonstrate that the profiles can still be accurately fit using multiple exponentials.
The steep dashed line shows a two-species \citet{m10} prediction. The more extended
dashed line shows the best-fit three-species model in which both the trap densities
and release times are allowed to vary. This analysis yields trap release times of
$\tau=\{0.74\pm 0.55,7.7\pm 4.3,37\pm 33\}\times 3212\mu$s with amplitudes of
$\{0.18\pm 0.10, 0.61\pm 0.3, 0.51\pm 0.26\}$ traps exposed to the 20,000~electron
charge cloud -- \ie\ a ratio of $1:3.38:2.85$. 

Pure E-centre complexes 0.44~eV below the conduction band are expected to
produce the next-longest trails, with $\tau\sim 180\times 3212\mu$s
\citep{jones00}. This is longer than measured, but the discrepancy may be due to
degeneracies in the fitting of decaying exponentials. This is a notoriously
difficult task because the exponentials become more and more similar to a
constant as they get longer. Our measurement of long trails is even more
difficult because the sky background in noisy and its subtraction is uncertain.
However, the degeneracy of exponentials also means that our trails can be
successfully fitted with different $\tau$ (and even an additional trap species)
by simply adjusting their normalisations. We therefore adopt the superior
three-trap model, with the fitted parameters after July 2006 and, assuming that
the third species are indeed pure E-centres, we update the pre-July 2006 values
to $\tau=\{0.48, 4.86, 20.6\}\times 3212\mu$s.


\subsection{Charge trap density} \label{sec:measurements}

%

\begin{figure*}
 \includegraphics[width=175mm]{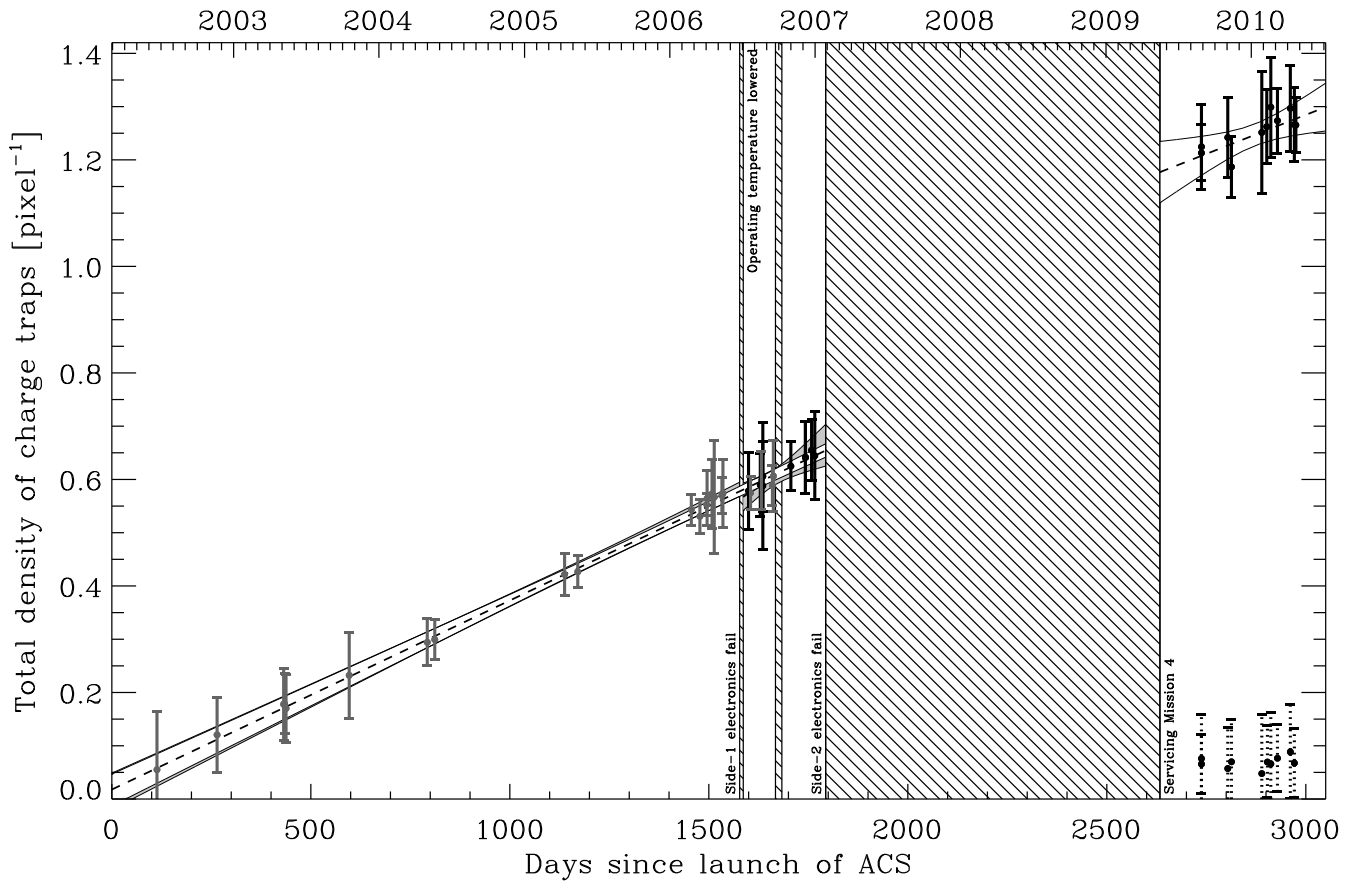}\vspace{-2mm}
 \caption{Measured density of charge traps in the ACS/WFC detectors, as they have accumulated over the lifetime of the camera. 
 Measurements assume three trap species in a ratio of $1:3.38:2.85$, with characteristic release times as described in the text.
 Grey (black) points indicate survey imaging acquired with a commanded gain setting of 1 (2), and all errors are $1\sigma$.
 Separate fits are shown to data before and after shuttle Servicing Mission 4, plus (noisier) fits to shorter periods in grey.
 Hatched regions indicate times when ACS was offline.
 Points with dotted error bars show the total absolute density of traps after correction.
 \label{fig:postsm4}\vspace{-3mm}}
\end{figure*}

As in \citet{m10}, we measure the effective trap density from the amount of
trailing in the first nine pixels behind warm pixels in archival HST imaging.
Large extragalactic surveys prove most useful to build up a uniform dataset
extending over a long period of time, and to isolate the warm pixels from
crowding by real astronomical sources. To span the entire lifetime of ACS, we
gather data from HST programmes GO-9075 and GO-10496 (PI: Saul Perlmutter), 
GO-9822 and GO-10092 (PI: Nick Scoville), GO-10896 (PI: Paul Kalas), GO-11563
(PI: Garth Illingworth) and GO-11600 (PI: Benjamin Weiner). The first four of
these were observed using a commanded gain setting of 1 and the last three with
a setting of 2; these are distinguished as grey and black data points
respectively in figure~\ref{fig:postsm4}. The exposure time and filters -- hence
the background level -- also vary between programmes.

Up to July 2006, we find that the total effective charge trap density increases
linearly over time\footnote{\citet{m10} demonstrated that monthly annealings do 
not reduce the effective trap density, so we also ignore them here.}. The trap
density extrapolates back to a value at launch of a remarkably low value of
$\rho_q=0.014\pm0.04$ per pixel. These manufacturing and process traps were
dominated within less than two months by radiation-induced traps, which were
created in orbit at a rate of $(3.60\pm0.26)\times 10^{-4}$ per day.


After July 2006, the traps themselves continued to accumulate at a rate of
$(4.77\pm2.76)\times 10^{-4}$ per day. Lowering the operating temperature and
lengthening the trails immediately reduced the amount of spurious flux in the first
nine pixels by 22\%. According to our model, however, the same amount of flux was
lost, but it was just moved further. The continuity of the apparent trap density
around this time, through changes in operating temperature, background level and
default gain settings, provide a strong vindication of our model. Overall, the total
trap density until January 2007 is well fit by
$\rho_q=(0.50\pm0.018)+(t-1359)\,(3.55\pm0.22)\times 10^{-4}$ per pixel, where $t$ is the number
of days after 1 March 2002.

Since January 2007, degradation has been more rapid. 
Because of the long period when ACS was offline, it is not clear exactly when this
damage accrued. It is quite possible that the radiation exposure simply increased. The
solar cycle maximum ended around 2006, and the density of charged particles in low
Earth orbit counterintuitively {\em increases} during solar minimum
\citep{sirianni06}. However, even though our data suggest the rate of trap creation
increased slightly in late 2006, it appears to have slowed again since 2009. A more
likely scenario is that the damage built up abruptly while ACS was offline and warm,
before shuttle Servicing Mission 4. The subsequent trap density is
best-fit by $\rho_q=(1.25\pm0.020)+(t-2873)\,(2.93\pm2.25)\times 10^{-4}/$pixel.



\subsection{Algorithmic development}

\cite{a10} and \cite{short10} ingeniously invoke a first-order symmetry of the
readout process to increase the speed of the readout (and correction) algorithm.
Electrons beginning 2048 pixels from the readout register undergo this many
pixel-to-pixel transfers during readout. Each time, electrons may be captured or
released by charge traps but, if the number of free electrons is high and the
density of traps is small, every transfer is statistically similar. The fast
algorithm  performs only one transfer for each cloud of electrons, then
multiplies its effect by the number of transfers it will see. This can be
implemented quickly in practice by sweeping one pixel's worth of traps up the
CCD (rather than all the electrons down the CCD). This approach is very
powerful: we confirm that it decreases runtime by a factor of $\sim1000$ and
even still allows for secondary charge capture, whereby a trailed electron can
be subsequently recaptured and retrailed. 

The limitation of this algorithm is that all of the capture (and recapture) of
electrons is implemented at the level appropriate to the size of the electron
cloud in the raw image. In the real readout process, as electrons are gradually
removed from an image peak, the electron cloud shrinks and fewer are
subsequently captured. Similarly, as electrons build up in a trail, the cloud
grows and becomes exposed to more traps. This effect is illustrated as the
difference between the dashed and solid lines in figure~\ref{fig:trail} and is
most severe for faint sources -- from which the fast algorithm even makes it
possible to trail more electrons than are available. We propose a compromise
between speed and accuracy by using one transfer to represent the first $n_t$
transfers, then performing a new transfer to represent the next $n_t$ and so on.
Thus the height of local maxima slowly reduces and the height of trails slowly
increases. We find that $n_t=140$ still provides a factor of $\sim$70 speedup,
while producing a trailed image within 1 electron of that produced by the full
algorithm everywhere on the detector for model parameters appropriate in early
2010.

Only one iteration of the \cite{bristow03im} algorithm \citep[see table~1
in][]{m10} was required to correct the circa 2004 COSMOS survey. This was
because the low density of charge traps implied only a small, perturbative
correction. To correct more recent data, \cite{a10} implemented five iterations.
To qualitatively justify the number of iterations, it is merely necessary to
test for convergence by calculating the difference to the corrected image after
each step. Typical science images from early 2010 change by only one electron in
a handful of pixels after three iterations, and by less than an electron in
every pixel after four. Since each iteration has a large price in run time, we
shall henceforth stop at the third iteration.

\section{IMAGE CORRECTION} \label{sec:implementation}

We use our updated CCD readout model to correct science imaging throughout the
lifetime of ACS, following the same procedure as \citet{bristow03im}. The points
with dotted error bars in figure~\ref{fig:postsm4} show the effective density of
charge traps after correction, which are a factor of 20 lower than in the raw
data and consistent with image quality in the first six months of operations.
For the sake of clarity, equivalent post-correction measurements are not shown
for earlier epochs, but these recover about the same factor of 10-15 correction
seen in \citet{m10}. Thus, ironically, as the CTI has got worse, the trailing
has become easier to measure and the correction has become more accurate!

\begin{figure}
 \begin{center}
 \includegraphics[width=41mm]{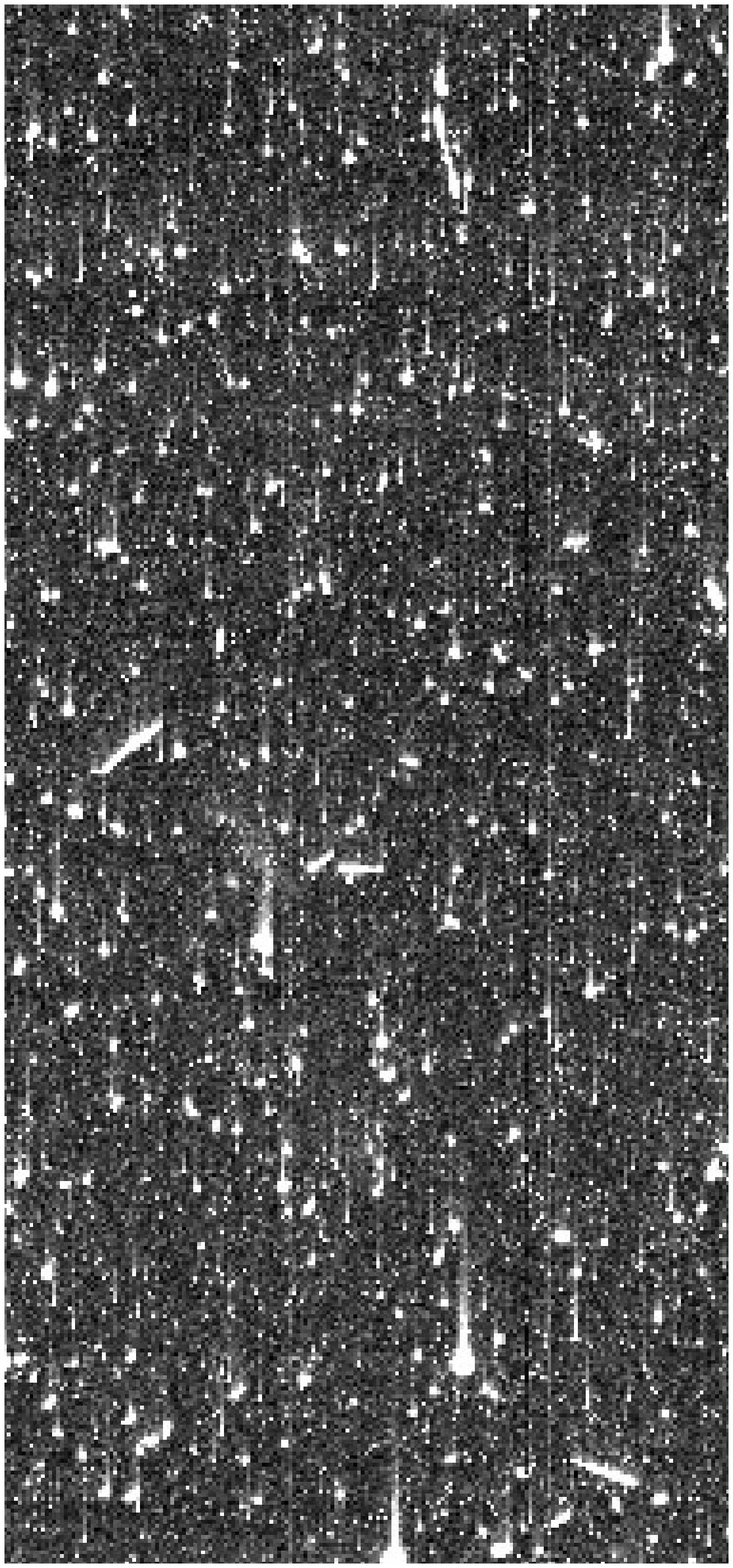}
 \includegraphics[width=41mm]{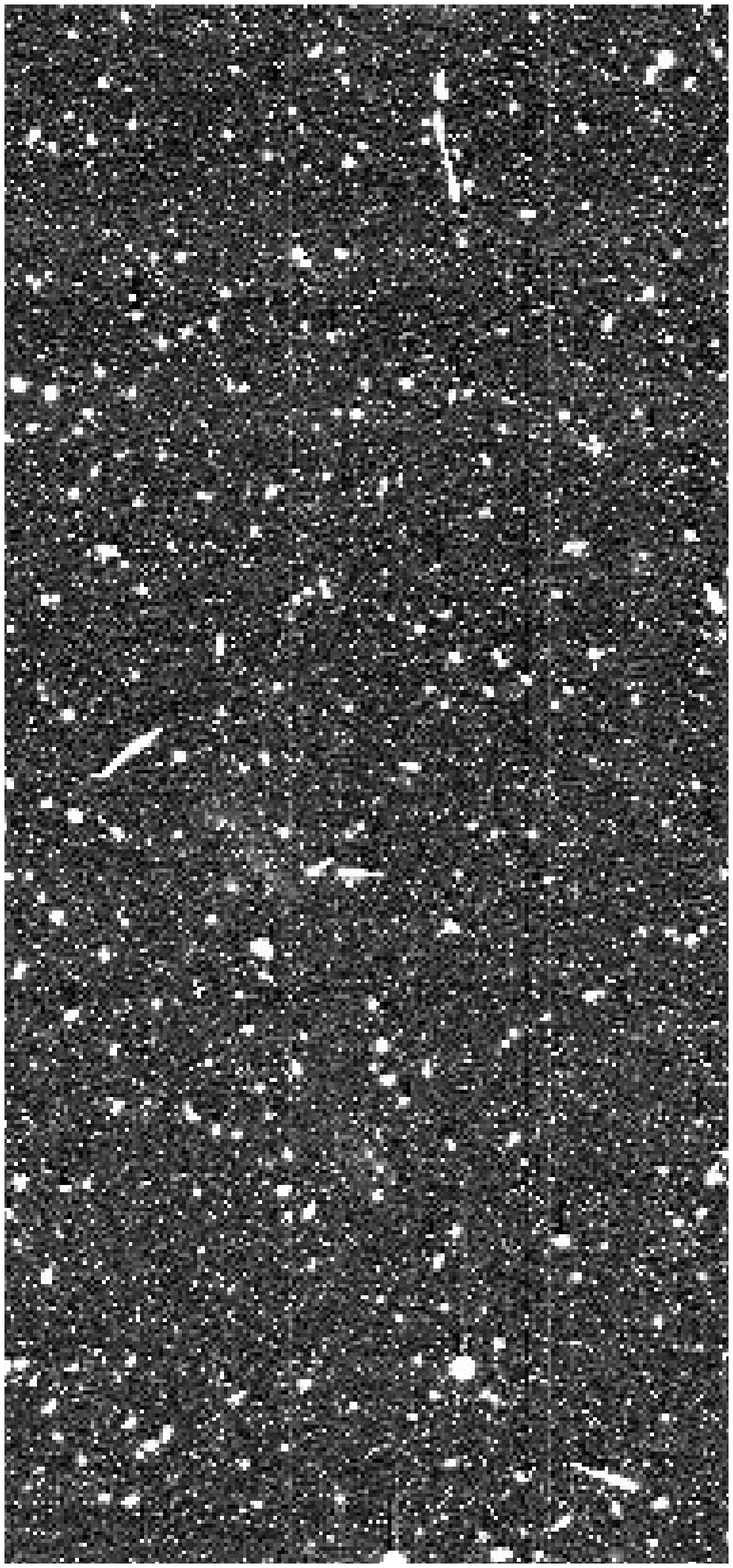}
 \end{center}
 \caption{A typical {\tt \_raw} ACS/WFC science exposure from 
 early 2010 (HST-GO-11689, PI: Renato Dupke) before (left) and after (right)
 CTI correction.
 The $380\times 820$ pixel area selected is furthest on the detector from the
 readout register, and the logarithmic colour scale is chosen intentionally to
 highlight the CTI trails.
 \label{fig:beforeafter}}
\end{figure}

Figure~\ref{fig:beforeafter} shows a region of a typical exposure, which was
intentionally {\em not} used when measuring parameters of the readout model. The
charge trailing that is now readily apparent in visual inspection of recent ACS
images is successfully removed by our correction scheme.



\section{Discussion} \label{sec:conclusions}

We have developed a physically-motivated model of the readout and Charge
Transfer Inefficiency (CTI) in the ACS/WFC detectors throughout their lifetime.
We find that there are approximately 1.3 charge traps per pixel in mid-2010,
split between three different species. The extended trails produced by these
traps can be accurately modelled as a sum of three decaying exponentials. We
also used our model to correct images, reducing the amount of trailing by a
factor of $\sim$20, to a level seen in the first six months of orbital
operations.
As with \citet{chiaberge}, we still find no evidence for significant
serial CTI (trailing perpendicular to the main trails, created by charge traps
in the serial readout register), and therefore ignore this effect.

When building our model, we adopted the best available measurements from science
imaging (which we performed) and dark exposures \citep[from][]{a10}. The dark
exposures were particularly useful to constrain the extended shape of trails out
to $\sim$100 pixels and thus provide better correction of object photometry
\citep{rhodes10}. Where measurements disagreed, the data's support for our
physical model encouraged us to first extrapolate measurements obtained in the
most reliable regime. 

Removing the final few percent of CTI trails might require detailed
investigation of such disagreements. In particular, there is mounting evidence
that trails behind sources of different flux may change in {\em shape} as well
as amplitude. A slight steepening of faint trail was also present in
\citet{m10}, but ascribed to uncertain background level and read noise. Read
noise is added to an image {\em after} CTI, so creates spurious faint peaks that
are not trailed, and act to spuriously steepen the true mean trail when they are
accidentally included in the average. A physical effect that we do not model,
but which might also affect faint trails, is the breakdown of the volume-driven
charge packet model at very low flux levels discussed by \cite{short10}.
However, while this is important in Time-Delay Integration (TDI) mode
observations (and potentially dark exposures), it is not so in science imaging
where a large zodiacal sky background is always present. If anything, the effect
would also predict shallower trails behind faint sources, from high-$\tau$
traps. A second physical mecahnism by which the trail could change shape could
be the onset of surface full well traps above a certain flux. However, this
explanation seems unlikely at a value of 20,000~electrons (\cf\  $>80,000$ full
well depth), and because such traps would have been present since manufacture,
while almost all appear to have accumulated over time at the same rate. We shall
therefore continue to use a single trail profile, but recommend further testing
of this apparent shape change, for example in combination with mean-variance
measurements at a range of flux levels to determine whether the shape change is
gradual or discreet and, if discreet, whether it coincides with other discontinuities.

The 4C decrease in the operating temperature of the ACS/WFC detectors in July 2006
did not affect the density of charge traps, or the amount of flux lost from a
source. Howver, it lengthened their release times and the amount of spurious flux
in the first 9 pixels behind a source fell by 22\%, which benefits some
astronomical measurements. Weak lensing measurements suffer by way of a spurious
shear signal induced the readout direction. Extrapolating from the trap
characterisation of \citet{rhodes10}, we estimate in mid-2010 a mean shear of
$\sim$5\% in galaxies detected at a S/N of $10$. Similarly, we expect a value
twice as bad for a galaxy at the chip gap (but zero at the edge), and about half
as bad in a galaxy one magnitude brighter. Verifying this in practice would
require a new survey similar in size to COSMOS \citep{scoville07}.

Most dramatically, the charge trap density increased $\sim80$\% more than
expected between the failure of ACS in January 2007 and its resumpion of
activities after shuttle Servicing Mission 4. It is not yet clear whether this
degradation is related to the decrease in operating temperature, the increase in
temperature while ACS was offline, or coincidentally due to the ending of the
solar cycle. Our  current analysis uses almost all the suitable archival data
currently available without yielding conclusive evidence. To resolve this issue,
we plan continued monitoring for a further year.

\section*{Acknowledgments}

I would like to thank the CTI working group at STScI, including Linda Smith, Ray
Lucas, Pey Lian Lim, Norman Grogin and David Golimowski. I am especially grateful
to Jay Anderson who shared his paper before publication. Roger Smith and Alex
Short provided physical insight, and Chris Stoughton provided code. Suggestions about
temperature effects from the referee proved invaluable: thank you!


The author is supported by STFC Advanced Fellowship \#PP/E006450/1 and European 
Research Council grant MIRG-CT-208994.
This work was based on observations with the NASA/ESA {\em Hubble Space
Telescope}, obtained at the Space Telescope Science Institute, which is operated
by AURA Inc, under NASA contract NAS 5-26555.
Data were used from programmes
GO-9075, GO-9822, GO-10092, GO-10496, GO-10896, GO-11563, GO-11600 and GO-11689.


\label{lastpage}

\end{document}